\documentclass[preprint2]{aastex}

\newcommand{\be}{\begin{equation}}
\newcommand{\ee}{\end{equation}}

\begin{document}

\title{Mitigating Charge Transfer Inefficiency in the Chandra X-ray
Observatory's ACIS Instrument}

\author{L. K. Townsley, P. S. Broos, G. P. Garmire, J. A. Nousek}
\affil{Department of Astronomy \& Astrophysics, The Pennsylvania State 
University, 525 Davey Lab, University Park, PA 16802}
\email{townsley, patb, garmire, nousek @astro.psu.edu}

\begin{abstract}
The ACIS front-illuminated CCDs onboard the Chandra X-ray Observatory
were damaged in the extreme environment of the Earth's radiation belts,
resulting in enhanced charge transfer inefficiency (CTI).  This
produces a row dependence in gain, event grade, and energy resolution.
We model the CTI as a function of input photon energy, including the
effects of de-trapping (charge trailing), shielding within an event
(charge in the leading pixels of the 3$\times$3 event island protect
the rest of the island by filling traps), and non-uniform spatial
distribution of traps.  This technique cannot fully recover the
degraded energy resolution, but it reduces the position dependence of
gain and grade distributions.  By correcting the grade distributions as
well as the event amplitudes, we can improve the instrument's quantum
efficiency.  We outline our model for CTI correction and discuss how
the corrector can improve astrophysical results derived from ACIS
data.

\end{abstract}



\section{Introduction
\label{sec:intro}}
	
The Advanced CCD Imaging Spectrometer (ACIS) instrument on the Chandra
X-ray Observatory \citep{odell98} employs bulk front-illuminated (FI)
and back-illuminated (BI) CCDs to give good spectral resolution and
good 0.2--10~keV quantum efficiency (QE) \citep{bautz98}.  These
devices, designed and manufactured at MIT's Lincoln Laboratories
\citep{burke97}, couple with the Chandra mirrors to provide
spatially-resolved X-ray spectroscopy on spatial scales comparable to
ground-based optical astronomy.

Due to the unanticipated forward scattering of charged particles
(probably mainly $\sim$100 keV protons, \cite{prigozhin00}) by the
Chandra mirrors onto the ACIS CCDs, the FI devices have suffered
degraded performance on-orbit, most pronounced at the top of the
devices, near the aimpoint of the imaging array.  This radiation first
encounters the gate structure of FI devices, causing charge traps that
increase the charge transfer inefficiency (CTI) of these devices,
resulting in gain variations, event grade distortions, and degraded
energy resolution as a function of row number.  Due to the
manufacturing process, the ACIS BI devices have always shown modest CTI
effects, but the silicon covering the gates protects them from severe
radiation damage on-orbit so they show no additional CTI.  Our ongoing
efforts to mitigate the BI CTI by modeling and post-processing the
data allow us to address the new problem of FI CTI in a
timely fashion.

\begin{figure}
\plotone{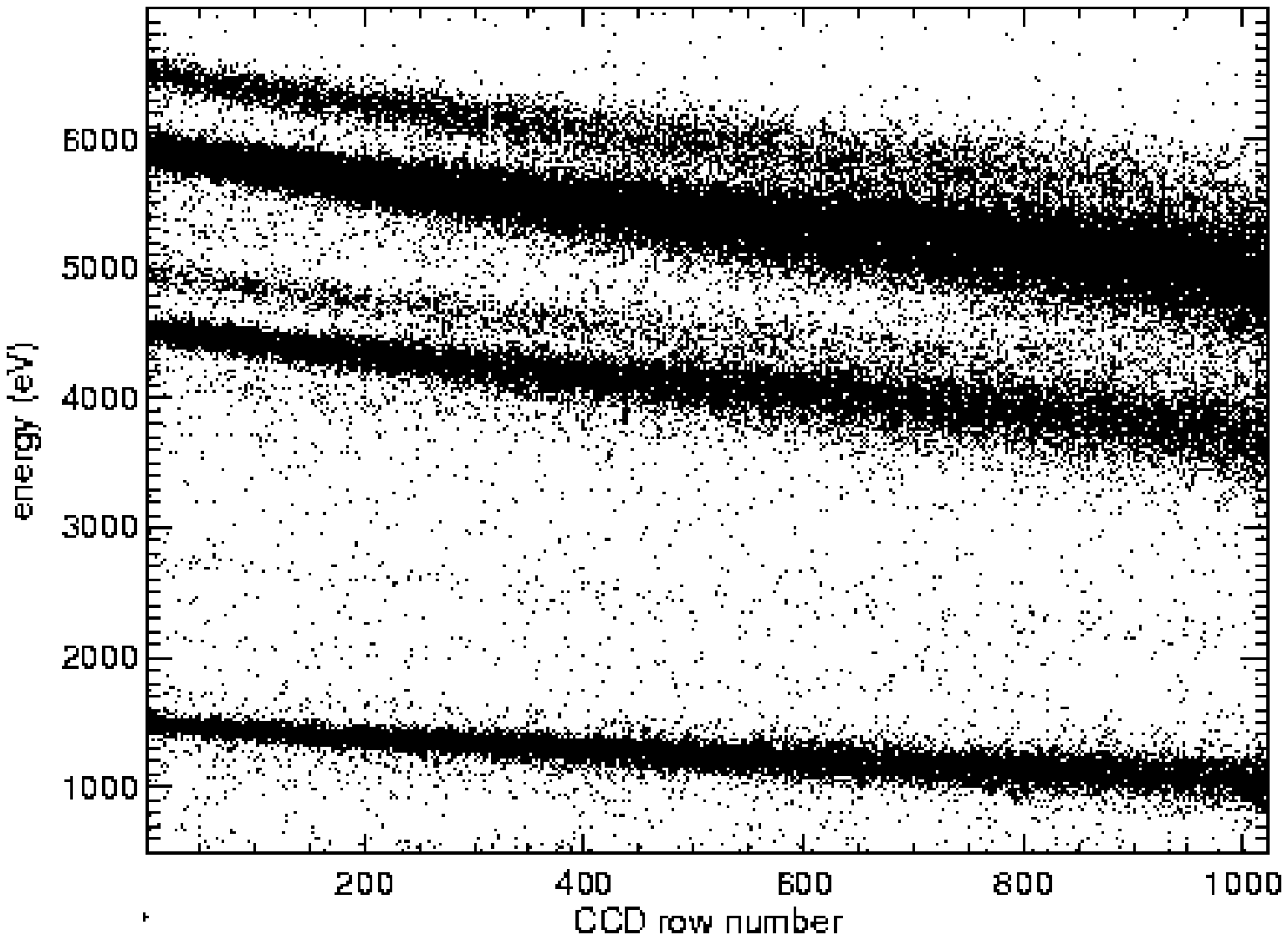}
\plotone{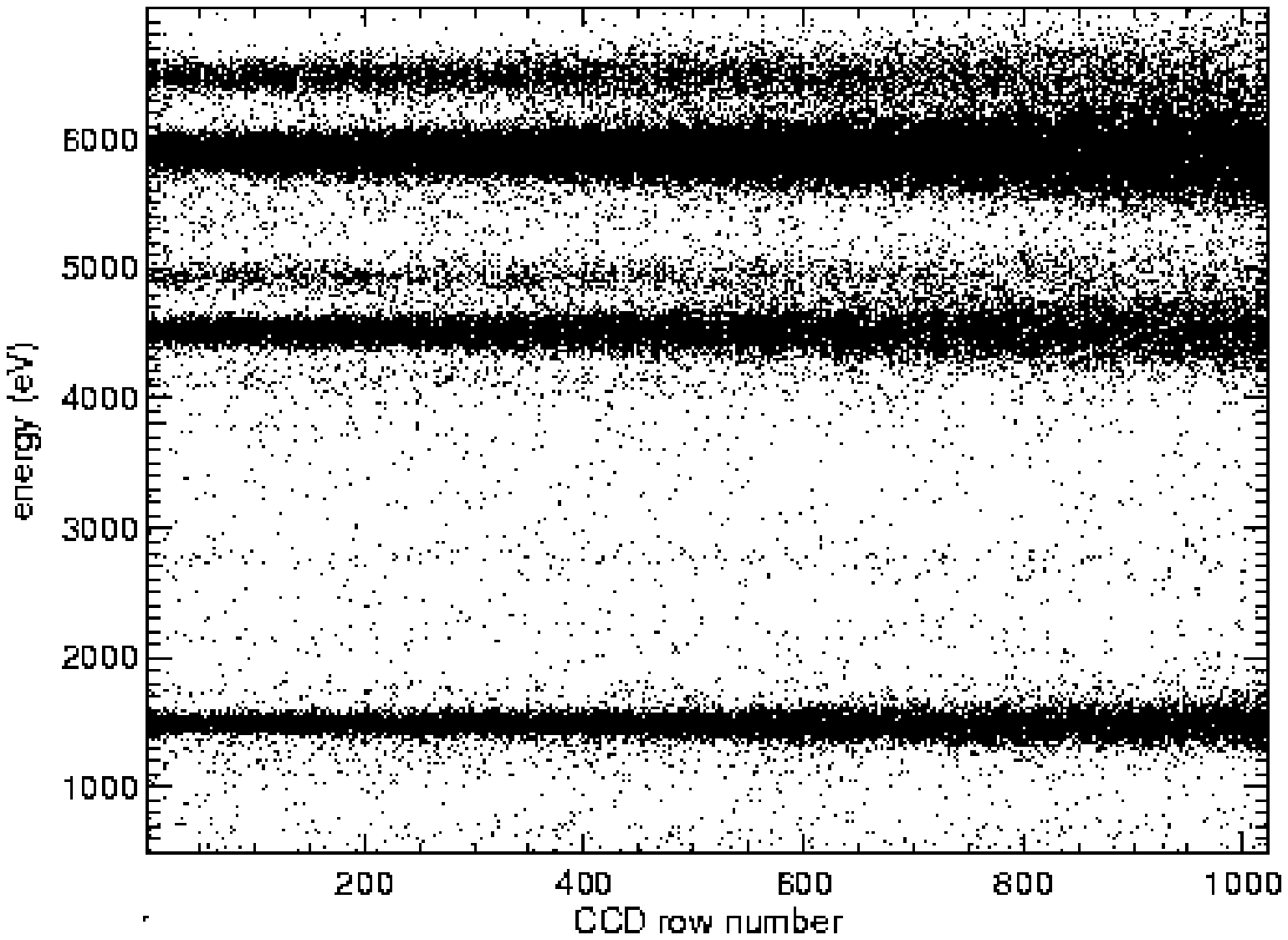}
\figcaption{Upper panel:  CTI in the I3 chip, illustrated with data from the ECS.  Note the row-dependent gain and
energy resolution and the energy-dependent charge loss.  Lower panel: CTI-corrected version of these data.
\label{fig:allenergy-cti}}
\end{figure}

Figure~\ref{fig:allenergy-cti} (upper panel) illustrates the effects of
CTI on the ACIS I3 chip by showing the row-dependent gain and energy
resolution for spectral lines in the External Calibration Source (ECS)
(from low to high energy:  Al~K$\alpha$, Ti~K$\alpha$,$\beta$,
Mn~K$\alpha$,$\beta$).  The data were filtered to keep ASCA-like grades
0, 2, 3, 4, and 6 (``g02346'', \cite{yamashita}) only.  Note that the
charge loss is more severe at higher energies (the loci steepen with
line energy).  This dataset, obtained after the damage was discovered
and satellite operations were modified to prevent further degradation,
is representative of the state of the CTI from mid-September 1999 to
the end of January 2000, when the ACIS focal plane temperature was
lowered from -110C to -120C.  At this new colder temperature, the trap
population that causes the row-dependent energy resolution is partly
suppressed \citep{gallagher98}.


CTI redistributes the charge in each event among the pixels in the
3$\times$3 island (grade ``morphing'') as some charge traps release
their charge on short timescales, so that charge reappears in adjacent
pixels.  This grade morphing, coupled with the standard g02346 grade
filtering, causes the QE to vary with event energy and position on the
device.

We have developed and optimized a model of CTI that accounts for charge
loss and the spatial redistribution of charge (``trailing''), in both
the parallel and serial registers, thus reproducing the
spatially-dependent gain, QE, and grade distribution of these devices.
By forward modeling of these effects separately for each event, we
produce a ``CTI-corrected'' event list that is useful for data analysis and astrophysical interpretation.

For the BI devices, we must model parallel and serial charge loss and
trailing independently to match the data.  There are charge traps in
the imaging array, framestore array, and serial register.  Since the FI
devices were damaged by radiation that penetrates only a few microns
into the device \citep{prigozhin00}, their (shielded) framestore arrays
and serial registers are protected.  So, although the CTI is more
pronounced on FI devices, it is easier to model than in the BI
devices.  Our techniques are described in more detail, with more
emphasis on the BI devices, in a separate article \citep{townsley00b}.


\section{The Model
\label{sec:model}}	
	
Our CTI model is phenomenological, characterizing the effects of CTI in
the data rather than directly modeling the spatial distribution and
time constants of the trap population (see \cite{gallagher98} and
\cite{krause99} for examples of such physical models). The charge lost
and trailed into the adjacent pixel are given by $L = k_{1}C^{k_{2}}$;
$T = k_{3} L^{k_{4}}$ where $C$ is the charge in a single pixel, $L$
is the charge lost per pixel transfer, $T$ is the charge trailed per
pixel transfer, and the $k$'s are the CTI parameters.  We chose power
law functional forms for these relations because they adequately fit
the data and ensure zero charge loss or trailing when there is zero
charge present.  The parameters $k_{1}$--$k_{4}$ are determined by
fitting lines to pixel amplitude versus x and y chip positions for many
energies.  Examples of each of these fits are shown in
\cite{townsley00b}.  To apply CTI to a given event, these equations are
calculated for each pixel in the event, then the resulting numbers are
multiplied by the number of pixel transfers to determine how much
charge to remove from each pixel in the event.  The trailed charge is
then added back into the appropriate adjacent pixel.

Note that our model is ``local'' -- prediction of the amount of charge
lost by a particular event does not consider the possibility that
another event falling in the same columns but closer to the amp will
have filled some portion of the traps, reducing the charge loss from
its nominal value \citep{gendreau95}.  We do however consider shielding
effects that may occur within the nine pixels of a single event.  For
example, an isolated pixel of height 1000 Data Numbers (DN, the digital
units of charge output by ACIS, $\simeq$4~eV) will encounter more empty
traps (suffer more loss) than a pixel of height 1000~DN that is
immediately following a pixel of height 400~DN.  The ``leading''
pixel's charge is sacrificed to fill the traps encountered by the
event, shielding the 1000~DN pixel.  Thus we apply our loss model only
to the portion of the pixel's charge that is larger than its
predecessor's.  In this example, the 1000~DN pixel will suffer loss and
trailing approximately the same as a 600~DN isolated pixel.  Details of
the shielding model are given in \cite{townsley00b}.

Although ACIS FI devices do not exhibit serial CTI, they do show a
small charge loss effect which is proportional to the distance between
the CCD pixel and the left or right edge of the CCD.  This is believed
to be caused by the slight drop in clocking voltages suffered between
the middle of the CCD and the supply leads at the chip edges
(M.~W.~Bautz, private communication).  CTI scales the effect so that
these gain variations are row-dependent, with the bottom rows of the
chip showing virtually no variation and the top rows varying by up to
$\sim$100 DN at high energies ($\sim$6 keV).  We model this gain
variation for each amp as a linear function of distance from the
readout node, averaging over all rows.

To account for nonlinearities not modeled directly by the fits
described above, we also incorporated a ``CTI deviation map'' into our
model that refines the CTI for each event's specific location on the
chip.  This map is produced by examining the residual gain variations
left after the basic model is applied.  By correcting ECS data with and
without this deviation map, we have determined that including it
improves the final results.  See \cite{townsley00b} for details.

\section{The Corrector  
\label{sec:corrector}}

Having modeled CTI, we can try to remove its effects from actual ACIS
events, taking a forward-modeling approach.  For each
{\em observed event}, we first hypothesize the nine pixel values of a
corresponding {\em true event}, {\em i.e.}\ the event that would have
been obtained if no CTI effects were present.  This {\em true event} is
passed through the CTI model, producing a {\em model event}.  The
hypothesis is then adjusted pixel-wise by adding the difference between
{\em observed event} and {\em model event} to {\em true event} and
repeating the process.  This iterative adjustment continues until {\em
model event} and {\em observed event} differ by less than one DN, at
which point {\em true event} constitutes the CTI-corrected event.

The performance of the corrector in removing the observed spatial
variation of event energy in ECS data is shown in the lower panel of 
Figure~\ref{fig:allenergy-cti}.  The corrector has largely removed
the gain changes across the device.  Since the spectral broadening with
row number is primarily the result of a random process involving one of
the trap species \citep{antunes93}, it cannot be suppressed in our
reconstructed events.

The success of the CTI corrector in regularizing the grade distribution
at the top of an FI device is shown in Table~\ref{tbl:grade_migration},
combining a broad range in energies (0.5--7 keV) made up of the Al, Ti,
and Mn calibration lines.  For example, 43.6\% of events migrated from
grade 2 to grade 0 during CTI correction.



\begin{deluxetable}{crrrrrrrr|rc}
\tabletypesize{\small}

\tablecaption{Correcting Grade Migration in ECS Data \label{tbl:grade_migration}}
\tablewidth{0pt}

\tablehead{ 
 & \multicolumn{8}{c}{\underline{original grade}} & \multicolumn{2}{c}{\underline{new distribution}}\\
\colhead{new grade} & \colhead{0} & \colhead{1} & \colhead{2} & \colhead{3} & \colhead{4} & \colhead{5} & \colhead{6} & \colhead{7} & \colhead{top} & \colhead{bottom}
}

\startdata
0 & 9.3 & 0.0 & 43.6 & 0.0 & 0.0 & 0.0 & 0.0 		& 0.0 	& 52.9 & 49.7 \\
1 & 0.0 & 0.0 & \underline{0.1} & 0.0 & 0.0 & 0.2 & 0.0 & 0.1 	& 0.4  & 0.3  \\
2 & 0.0 & 0.0 & 11.3 & 0.0 & 0.0 & 0.0 & 0.0 		& 0.0	& 11.3 & 16.0 \\
3 & 0.0 & 0.0 & 0.0 & 0.3 & 0.3 & \textbf{0.2} & 4.7 	& 0.0	& 5.5  & 6.0  \\
4 & 0.0 & 0.0 & 0.0 & 0.3 & 0.3 & \textbf{0.1} & 4.6 	& 0.0	& 5.3  & 5.9  \\
5 & 0.0 & 0.0 & 0.0 & 0.0 & 0.0 & 0.3 & \underline{0.3} & 0.3	& 0.9  & 0.5  \\
6 & 0.0 & 0.0 & 0.0 & 0.0 & 0.0 & 0.0 & 7.2 	& \textbf{4.1}	& 11.3 & 13.5 \\
7 & 0.0 & 0.0 & 0.0 & 0.0 & 0.0 & 0.0 & 0.0 		& 12.3 	& 12.3 & 8.0  
\enddata

\tablecomments{For events falling at the top of the detector (rows 824:1024) 
where CTI is most severe, Columns 2--9 show the percentage of events
having a certain ASCA grade before correction (table column) and a
certain ASCA grade after correction (table row).  Entries in
boldface represent events gained by CTI correction; entries
underlined represent events lost, assuming the standard g02346 grade
filtering.  Column 10 shows the corrected grade distribution at the top
of the detector.  For comparison column 11 shows the corrected grade
distribution at the bottom of the detector (rows 1:201).  Summing down
Columns 2--9 would give the grade distribution at the top prior to
correction. 
}

\end{deluxetable}

\begin{figure}
\plotone{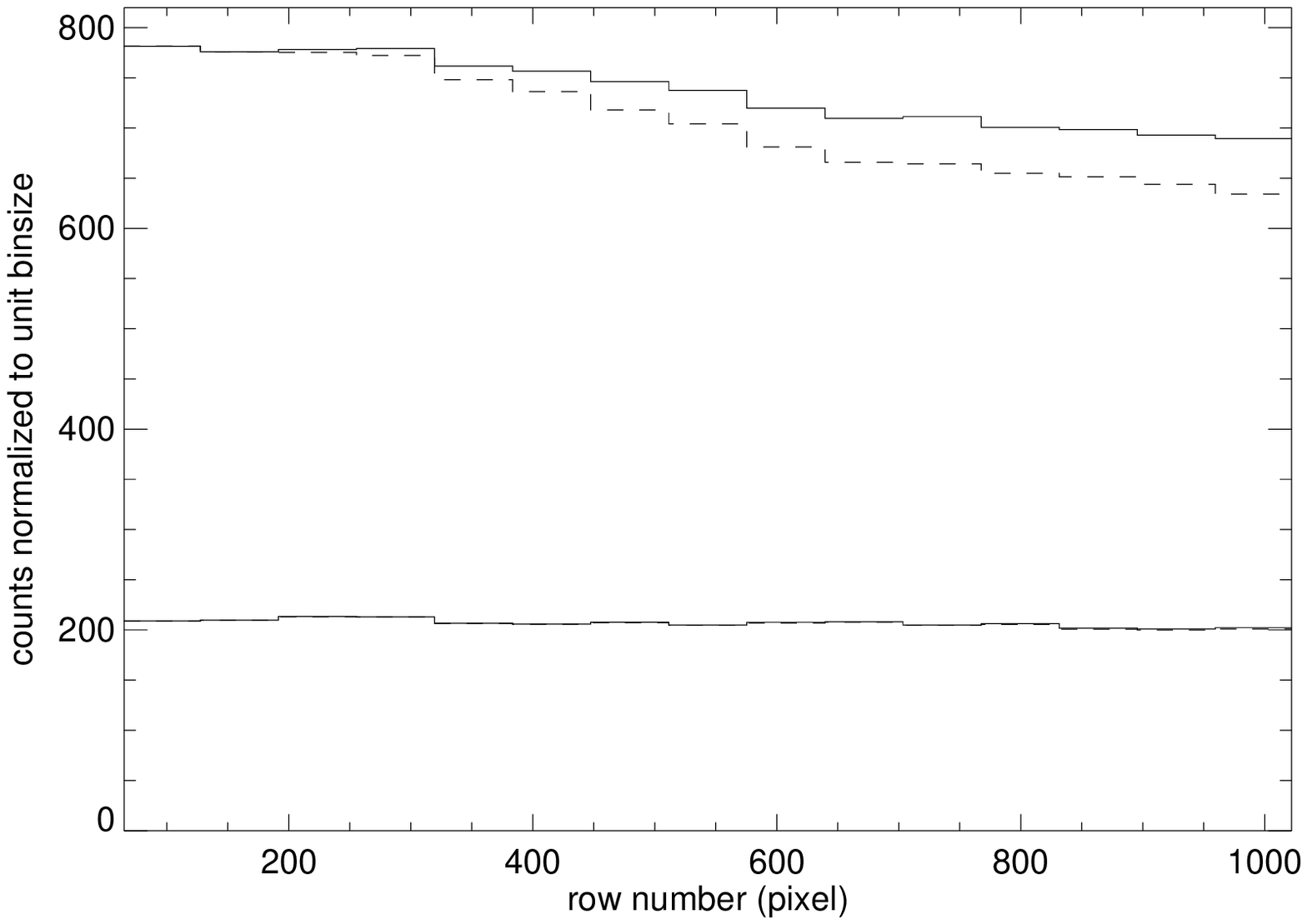}
\figcaption{Relative QE at Al~K (bottom) and Mn~K (top) across the I3 device,
before (dashed lines) and after (solid lines) CTI correction, ASCA
g02346 only.  The binsize is 32 rows.  For Al~K, corrected and uncorrected
QE are nearly identical.  Uncorrected Mn K data show $\sim$19\% 
fewer events at the top of the device than at the bottom; corrected data
show a modest QE improvement but still $\sim$11\% fewer events at the
top of the device compared to the bottom. \label{fig:qe}}
\end{figure}

Table~\ref{tbl:grade_migration} demonstrates an improvement in QE after
CTI correction.  Figure~\ref{fig:qe} illustrates this in more detail.
For low-energy events ({\em e.g.}\ Al K) the charge is more spatially
concentrated, leading to more single-pixel events.  CTI causes these to
morph into grades that are still captured in the standard grade
filtering, so the QE at the top of the device is not affected by CTI.
At higher energies ({\em e.g.}\ Mn K) events are often intrinsically
split, so grade morphing causes them to be lost to ASCA-like grade 7.
The corrector can recover some of these events, but clearly the QE is
still row-dependent.  This is due in part to telemetry constraints:
certain ACIS grades are not transmitted to the ground because they are
primarily cosmic ray events and would saturate telemetry.  When CTI
causes legitimate photon events to grade morph into these grades, they
are lost forever.  Other events are not recovered into their original
grades because the corrector is unable to recover charge that has been
eroded by CTI down below the split threshold -- at that level, the
charge is indistinguishable from noise and the corrector is purposely
not allowed to include such pixels in its reconstruction.  Note that
the bottom quarter of the device is largely unaffected by the CTI
degradation and thus retains the original performance of the detector.

The CTI corrector consists of a set of IDL (Research Systems Inc.)
programs and parameter files that instantiate the model for each
amplifier on each CCD, for a given epoch of the Chandra mission.  The
code is available (www.astro.psu.edu/users/townsley/cti/corrector/) and
may be of interest to other groups concerned with the results of
radiation damage on X-ray and optical CCD detector systems.  This
website is intended to provide {\em an example} of the corrector, not
an exhaustive and complete set of correction code immediately
applicable to any dataset.

\section{Model Uncertainties  
\label{sec:uncertainties}}

The FI corrector must be optimized to match the CTI present at the time
of the observation in question, since the CTI has changed over the
course of the mission.  This can be achieved by tuning the corrector's
parameters on the ECS dataset taken most closely in time to the target
observation.  There is an uncertainty introduced by using calibration
data to tune this or any other model:  we must assume that the CTI
measured by the ECS is representative of that present in celestial
data.  Most ACIS events are due to particle interactions and are not
even telemetered.  It is primarily these events that fill traps and
thus regulate CTI.  Since most of these particle events are from cosmic
rays that are not stopped by the Observatory structure, it is
reasonable to assume that ECS and celestial data have comparable
particle fluxes, thus a comparable trap population.

We have used the instrumental Au~L$\alpha$ (9.7 keV) and Ni~K$\alpha$
(7.5 keV) lines as rough diagnostics of the corrector's performance on
celestial data.  Tests show that the corrector works fairly well even
at these high energies (well above the energies used to tune it),
reducing the charge loss per pixel transfer by a factor of $\sim$10.
Better results might be obtained by hybridizing the corrector
optimization process to include both celestial and ECS data. 

Another implication of tuning the corrector to ECS data is that we have
no calibration information below Al~K$\alpha$ (1.486~keV).
Understanding the behavior of the CCDs between 0.2 and 1.5 keV is
important for obtaining accurate hydrogen column densities, comparing
ACIS results to ROSAT results, and inferring valid astrophysics for the
large number of soft sources Chandra will examine.  As a preliminary
example, we consider the supernova remnant E0102-72.3, used as a soft
calibration source for Chandra.  The CXC provided us with two
calibration datasets on this target (N. Schulz, private communication),
one with the target near the readout of Node 2 on the I3 chip (``obsid
49'') and a second with the target half-way up the chip (``obsid 48''),
on the same node.  An observation with the target near the top of the
chip is scheduled and will be included in subsequent analysis.

These data were obtained in January 2000 but CTI-corrected with the
code tuned to ECS data obtained the previous September.  Even though
the corrector was not tuned below 1.486 keV and was based on the CTI
present 4 months prior to the E0102-72.3 observations, the corrected
data for {\em both} observations showed the soft lines in this source
at the correct energies to within $\pm$1\% above 800 eV and to within
$\pm$2\% between 500 and 800 eV.  The grade distributions before and
after correction were identical, to within statistical uncertainties,
for obsid 49.  Grade migration was noticeable in the raw data from
obsid 48.  CTI correction restored the grade distribution to values
consistent with obsid 49.  With future data we hope that E0102-72.3 can
be used to calibrate the corrector as well as to check it.

The uncertainties outlined above should encourage ACIS users to treat
their data with care and caution.  This is especially important for
public calibration targets observed before mid-September 1999, as the CTI
was changing rapidly then.  


\section{Astrophysical Implications}

Users of ACIS data must be aware of CTI and its spatial and spectral
effects in order to generate astrophysically relevant results.  Without
CTI correction, accurate analysis of ACIS data requires a strongly
position-dependent response matrix (``.rmf file'') and QE map (part of
the ``.arf file'').  These of course depend on the user's choice of
grade filtering; it is not clear that the usual ASCA-like grade
filtering is optimal for sources far from the readout nodes.

Our CTI corrector improves the photon energy estimates and
significantly corrects for grade morphing.  The corrector also improves
the energy resolution of the ACIS devices more than can be achieved by
only accounting for the row-dependent gain variations.  Using these
features, we recover linewidths in the External Calibration Source data
that are on average 10\% narrower (for both FI and BI detectors) than
parallel gain detrending alone can provide.

These features mean that CTI-correcting ACIS event lists will yield
several benefits for the end user.  Correcting for grade morphing
yields more high-energy counts and thus gives deeper source detection
for hard sources; since most source-finding algorithms rely on a
threshold number of counts to constitute a detection, adding even a
modest number of counts can increase the number of detections
significantly.  For example, $\sim$15\% more hard (2--8 keV) sources
were detected in the ACIS GTO observations of the Hubble Deep Field and
surroundings after CTI correction \citep{hornschemeier00}.  Corrected
event energies are necessary to obtain valid hardness ratios for faint
point sources and to compare the spectral properties of faint sources
distributed across a field.  Improved energies and more uniform QE will
also yield better spectrally-resolved imaging of extended sources, an
important technique for studying supernova remnants, for example.

The corrector cannot totally eliminate the position-dependent energy
resolution of the FI devices or the energy- and position-dependent QE.
Thus position-dependent response matrices and effective areas are still
necessary for the most accurate spectral fitting and flux
determinations.

The CXC provides a tool (``MKRMF'') that allows the user to generate
spatially-dependent .rmf files for spectral analysis of CTI-affected
data.  In principle, we might use this tool for ACIS spectroscopy while
still benefiting from the CTI corrector.  As an example, the event list
should be CTI-corrected before any grade filtering is applied, since
the corrector can ameliorate grade morphing.  Then the standard g02346
grade filtering will yield an event list containing valid X-ray events
that is larger than that obtained by filtering on the original event
list.  Selecting these events out of the original event list will give
the best subset of events for spectral analysis.  The MKRMF tool is
then still applicable, as the observer will be using the original event
list.  Such a method might be useful for spectral analysis of
moderately faint sources, where including additional photons would
improve the spectral fits. 

Also necessary for spectral fitting is the ``MKARF'' tool that
incorporates the CCD QE into the .arf files.  Currently, MKARF
overestimates the high-energy QE for FI devices, since it presumes
pre-launch values.  This will lead to underestimates of the high-energy
flux, corrupting the temperatures or spectral indices obtained in model
fits.  It should be possible, though, to treat CTI-induced QE changes
as an energy-dependent multiplicative correction to the .arf file after
it is generated, for either original or CTI-corrected data.

For even more detailed spectro-spatial analysis, especially for extended
sources, we have incorporated our CTI model into the Monte Carlo
simulator that we developed to model ACIS devices \citep{townsley00a}.
Work is ongoing (G.\ Chartas, private communication) to couple
this CCD model with astrophysical models from XSPEC \citep{arnaud96}
and raytracing via MARX \citep{wise97}.  This will eventually allow
complete forward modeling to reproduce the data and optimize spectral
fitting results.

Efforts continue throughout the entire Chandra community to improve the
performance of the ACIS devices and our understanding of that
performance.  We expect the methods described here to be useful for
current data but eventually to be superceded for future data by
modifications to the instrument operations and software that are under
development now at MIT.  Laboratory work by MIT \citep{prigozhin00} and
PSU \citep{hill00} scientists is yielding new insight into the damage
mechanisms and providing data useful for calibrating and assessing
mitigation techniques.  In the meantime, our CXC colleagues are
providing tools for coping with CTI-affected data and suggesting
observing strategies that allow most ACIS users to accomplish their
basic science goals.


\acknowledgments

Financial support for this effort was provided by NASA contract
NAS8-38252.  We thank our MIT/ACIS colleagues, especially Mark Bautz
and Gregory Prigozhin, for many helpful discussions and suggestions
regarding CCD modeling and for their lucid explanations of the device
physics of radiation damage and charge traps.  We thank the CXC and
MSFC Project Science Team for providing a forum for discussion of CTI
and other ACIS calibration issues and we especially thank Harvey
Tananbaum for a careful review of this paper that greatly improved its
relevance to the X-ray astronomy community.





\end{document}